\def\icalp{0}
\ifnum\icalp=0
\documentclass[11pt]{article}
\usepackage{fullpage}
\else
\documentclass{llncs}
\fi

\usepackage{tikz}
\usetikzlibrary{shapes,snakes}
\usepackage[ruled,noline]{algorithm2e}
\usepackage{graphicx,amsfonts,amsmath,amssymb,epsfig,hyperref,color}

\renewcommand{\paragraph}[1]{{\protect\vspace{8pt}\noindent\sc{#1}}}

\newlength{\saveparindent}
\newlength{\saveparskip}
\newenvironment{newitemize}{%
\begin{list}{$\bullet$}{\labelwidth=19pt%
\labelsep=7pt \leftmargin=12pt \topsep=3pt%
\setlength{\listparindent}{\saveparindent}%
\setlength{\parsep}{\saveparskip}%
\setlength{\itemsep}{3pt} }}{\end{list}}

\ifnum\icalp=0
\fi

\newcommand{\BE}{\begin{enumerate}} \newcommand{\EE}{\end{enumerate}}
\newcommand{\BI}{\begin{itemize}} \newcommand{\EI}{\end{itemize}}
\newcommand{\BDes}{\begin{description}}\newcommand{\EDes}{\end{description}}
\newtheorem{alg}{Algorithm}
\newcommand{\BA}{\begin{alg}} \newcommand{\EA}{\end{alg}}
\newtheorem{thm}{Theorem}
\newcommand{\BT}{\begin{thm}} \newcommand{\ET}{\end{thm}}

\newtheorem{lem}{Lemma}      
\newcommand{\BL}{\begin{lem}} \newcommand{\EL}{\end{lem}}

\newtheorem{clm}[lem]{Claim}
\newcommand{\BCM}{\begin{clm}} \newcommand{\ECM}{\end{clm}}

\newtheorem{fct}[lem]{Fact}
\newcommand{\BF}{\begin{fct}} \newcommand{\EF}{\end{fct}}

\newtheorem{techcor}[thm]{Corollary}
\newcommand{\BCo}{\begin{techcor}} \newcommand{\ECo}{\end{techcor}}

\newtheorem{cor}[thm]{Corollary}      
\newcommand{\BC}{\begin{cor}} \newcommand{\EC}{\end{cor}}
\newtheorem{prop}[thm]{Proposition}     
\newcommand{\BP}{\begin{prop}} \newcommand {\EP}{\end{prop}}
\newtheorem{conj} {Conjecture}      
\newcommand{\BCJ}{\begin{conj}} \newcommand{\ECJ}{\end{conj}}

\newtheorem{defn}{Definition}         
\newcommand{\BD}{\begin{defn}} \newcommand{\ED}{\end{defn}}

\def\FullBox{\hbox{\vrule width 8pt height 8pt depth 0pt}}
\newcommand{\ourqed}{\;\;\;\FullBox}
\newenvironment{ourproof}{\noindent{\bf Proof:~~}}{\(\ourqed\)}
\newcommand{\OBPF}{\begin{ourproof}} \newcommand {\OEPF}{\end{ourproof}}
\newenvironment{proofof}[1]{\noindent{\bf Proof of {#1}:~~}}{\(\qed\)}
\newcommand{\BPFOF}{\begin{proofof}} \newcommand {\EPFOF}{\end{proofof}}
\newcommand{\qedsketch}{\;\;\;\Box}

\newenvironment{smallproof}{\noindent{\bf Proof:~~}}{\(\qedsketch\)}
\newcommand{\bpf}{\begin{smallproof}} \newcommand{\epf}{\end{smallproof}}

\newcommand{\BEQ}{\begin{equation}} \newcommand{\EEQ}{\end{equation}}
\newcommand{\BEQN}{\begin{eqnarray}}\newcommand{\EEQN}{\end{eqnarray}}

\newcommand{\eqdef}{\stackrel{\rm def}{=}}

\renewcommand{\Pr}{{\rm Pr}}

\newcommand{\poly}{{\rm poly}}

\newcommand{\eps}{\epsilon}

\newcommand{\calP}{{\cal P}}

\newcommand{\E}{{\rm E}}

\newcommand{\todo}[1]{\iffalse {#1} \fi }

\ifnum\icalp=1
\fi
\begin{document}
\ifnum\icalp=1
\title{A Quasi-Polynomial Time Partition Oracle for Graphs with an Excluded Minor}
\else

\title{A Quasi-Polynomial Time Partition Oracle \\ for Graphs with an Excluded Minor}
\fi
\ifnum\icalp=1
\author{Reut Levi\inst{1} \and Dana Ron\inst{2}
}
\authorrunning{Reut Levi and Dana Ron} 
%
\tocauthor{Reut Levi and Dana Ron}
\institute{School of Computer Science, Tel Aviv University.
Tel Aviv 69978, Israel.
\email{reuti.levi@gmail.com},
\and
School of Electrical Engineering, Tel Aviv University.
Tel Aviv 69978, Israel.
\email{danar@eng.tau.ac.il}}
\else
\author{Reut Levi
\thanks{School of Computer Science, Tel Aviv University.
Tel Aviv 69978, Israel.
E-mail: {\tt reuti.levi@gmail.com}.
}
\and
Dana Ron
\thanks {School of Electrical Engineering, Tel Aviv University.
Tel Aviv 69978, Israel.
E-mail: {\tt danar@eng.tau.ac.il}.
}
}
\fi
\maketitle
\begin{abstract}
Motivated by the problem of testing planarity and related properties,
we study the problem of designing efficient {\em partition oracles\/}.
 A {\em partition oracle\/} is a procedure that, given access to the incidence lists representation
 of a bounded-degree graph $G= (V,E)$ and a parameter $\eps$, when queried on a vertex $v\in V$, 
 returns the part (subset of vertices) which $v$ belongs to
 in a partition of all graph vertices. The partition should be such that all parts are small,
 each part is connected, and if the graph has certain properties, the total number of
edges between parts is at most $\eps |V|$.
In this work we give a partition oracle for graphs with excluded minors whose query complexity is
quasi-polynomial in $1/\eps$, thus improving on the result of
Hassidim et al. ({\em Proceedings of FOCS 2009}) who gave a partition oracle with query complexity
exponential in $1/\eps$. This improvement implies corresponding
improvements in the complexity of testing planarity and other properties that are characterized 
by excluded minors as well as sublinear-time approximation algorithms that work under the promise
that the graph has an excluded minor.
\end{abstract}

\section{Introduction}

An important and well studied family of graphs is the family of {\em Planar Graphs\/}.
A natural problem is that of deciding
whether a given graph $G = (V,E)$ is planar. Indeed, there is variety of linear-time algorithms for
\ifnum\icalp=0
deciding planarity (e.g.~\cite{HT74,ET76,SH99,BM04,dFOR06}). 
\else
deciding planarity (e.g.~\cite{HT74,BM04}). 
\fi
However, what if one is willing to relax the decision task while requiring that
the algorithm be much more efficient, and run in {\em sub-linear\/} time?
Namely, here we refer to the notion of {\em Property Testing\/} where
the goal is to decide (with high success probability) whether a 
graph has the property (planarity)
or is  far from having the property (in the sense that relatively many edges-modifications
are required in order to obtain the property). 
Such a task should be performed by accessing only small portions of the input graph.

Another type of problem related to planar graphs is that of solving a certain
decision, search, or optimization problem, under the {\em promise\/} that the 
input graph is planar, where the problem may be hard in general. 
In some cases the problem remains hard even
under the promise (e.g., Minimum Vertex-Cover~\cite{GJS76}), while in other cases
the promise can be exploited to give more efficient algorithms than are
known for general graphs (e.g., graph Isomorphism~\cite{HW74}).
Here too we may seek even more efficient, sublinear-time, algorithms,
which are allowed to output approximate solutions. 

\sloppy
The problem of testing planarity, 
and, more generally, testing any minor-closed property of 
graphs\footnote{For a fixed graph $H$, 
$H$ is a {\em minor\/} of $G$ if $H$ is isomorphic to a graph that can be obtained 
by zero or more edge contractions on a subgraph of $G$. 
We say that a graph $G$ is {\em $H$-minor free\/} (or {\em excludes $H$
as a minor\/}) if $H$ is not a minor of $G$. A property $\calP$ (class of graphs) is
{\em minor-closed\/} if every minor of a graph in $\calP$ is also in $\calP$. 
Any minor-closed property can be characterized by a finite
family of excluded minors~\cite{RS:20}.}
was first studied by Benjamini, Schramm and Shapira~\cite{BSS08}. They gave
a testing algorithm whose query complexity and running time
are {\em independent\/} 
of $|V|$.\footnote{Their algorithm has two-sided error.
If one-sided error is desired, then 
for any fixed $H$ that contains a simple cycle,
the query complexity of one-sided error
\ifnum\icalp=1
testing of $H$-minor freeness is $\Omega(\sqrt{|V|})$~\cite{CGRSSS}.
\else
testing of $H$-minor freeness is $\Omega(\sqrt{|V|})$~\cite{CGRSSS}.
On the positive side, if $H$ is cycle-free, then there is a one-sided error
algorithm whose complexity does not depend on $|V|$~\cite{CGRSSS}.
\fi
}
This result was later improved (in terms of the dependence on the 
distance parameter, $\eps$) by Hassidim et al.~\cite{HKNO09}, who 
also considered sublinear-time approximation algorithms
that work under the promise that the graph has an excluded (constant size) minor (or more generally, for 
hyperfinite graphs as we explain subsequently).
They show how to approximate the size of the minimum vertex cover, 
the minimum dominating set and the maximum independent set of such graphs, to within
an additive term of $\eps |V|$ 
in time that depends only on $\eps$ and the degree bound, $d$, but not on $|V|$.

The main tool introduced by Hassidim et al.~\cite{HKNO09} for
performing these tasks is {\em Partition Oracles\/}.
Given query access to the incidence-lists representation of
a graph, a partition oracle provides access to a partition of the vertices 
into small connected components. 
A partition oracle is defined with respect to a class of graphs, $\mathcal{C}$,
and may be randomized. 
If the input graph belongs to $\mathcal{C}$, then with high probability
 the partition determined by the oracle is such that the number of edges 
between vertices in different parts of the partition is relatively small
(i.e., at most $\eps |V|$).
Such a bound on the number of edges between parts together with the bound on
the size of each part lends itself to designing efficient testing algorithms
and other sublinear approximation algorithms.

Hassidim et al.~\cite{HKNO09} provide a partition oracle for hyperfinite~\cite{Ele06} classes of graphs that
makes $2^{d^{\poly(1/\eps)}}$ queries to the graph, where $d$ is and upper bound on the degree. 
A graph $G = (V, E)$ is {\em $(\eps, k)$-hyperfinite} if it is possible to remove at 
most $\eps |V|$ edges of the graph so
that the remaining graph has connected components of size at most $k$.
A graph $G$ is {\em $\rho$-hyperfinite} 
for $\rho : \mathbb{R}_+ \to \mathbb{R}_+$ 
if for every $\eps \in (0,1]$, $G$ is $(\eps, \rho(\eps))$-hyperfinite.   
For graphs with an excluded minor,
(a special case of hyperfinite graphs),
they provide a partition oracle with query complexity  $d^{\poly(1/\eps)}$ 
(as detailed in~\cite[Sec.~2]{OPhd}).
Hassidim et al.~\cite{HKNO09} leave as an 
 open problem whether it is possible to design 
 a partition oracle for graphs with an excluded minor 
that has query complexity polynomial in  $1/\eps$.
In particular, this would imply an algorithm for testing planarity
whose complexity is polynomial in  $1/\eps$.

\subsection{Our Contribution}
In this work we present a partition oracle 
for  graphs with an excluded minor whose query complexity and running time
are  $(d/\eps)^{O(\log (1/\eps))} = d^{O(\log^2(1/\eps))}$, that is, quasi-polynomial in $1/\eps$.

\paragraph{Implications.} 
Hassidim et al.~\cite{HKNO09} show how it is possible to reduce the problem of testing $H$-minor freeness (for
a fixed graph $H$) to the problem of designing a partition oracle for $H$-minor free graphs.
Using this reduction they obtain a testing algorithm for $H$-minor freeness 
\ifnum\icalp=0
(and more generally,
for any minor-closed property, since such properties can be characterized by a finite
family of excluded minors~\cite{RS:20}) 
\else
(and more generally, for any minor-closed property)
\fi
whose query complexity and running time are $2^{\poly(1/\eps)}$.
As noted previously, this improves on the testing algorithm of Benjamini et al.~\cite{BSS08} 
for minor-closed properties, whose complexity is $2^{2^{2^{\poly(1/\eps)}}}$.
Using our partition oracle (and the reduction in~\cite{HKNO09}) we get
a testing algorithm whose complexity is $2^{O(\log^2(1/\eps))}$.

Other applications of a partition oracle for a class of graphs $\mathcal{C}$
 are constant time algorithms that work under the promise that 
the input graph belongs to $\mathcal{C}$, where in our case $\mathcal{C}$
is any class of graphs with an excluded minor. 
Under this promise, Hassidim et al.~\cite{HKNO09}  provide constant time 
$\eps |V| $-additive-approximation algorithms for the size of a minimum vertex cover, 
minimum dominating set and maximum independent set.
They also obtain an $\eps$-additive-approximation 
algorithm for the distance from not having any graph 
in $\mathcal{H}$ as an induced subgraph where $\mathcal{H}$ is a fixed subset of graphs. 
Combined with our partition oracle, the query complexity of these algorithms drops from 
$d^{\poly(1/\eps)}$ to $(d/\eps)^{O(\log (1/\eps))} = d^{O(\log^2(1/\eps))}$. 

\paragraph{Techniques.}
As in~\cite{HKNO09}, our partition oracle runs a local emulation of a global
partitioning algorithm. Hence, we first give a high-level idea of the global partitioning
algorithm, and then discuss the local emulation. Our global partitioning algorithm
is based on the global partitioning algorithm of~\cite{HKNO09}
for graphs with an excluded minor, as described in~\cite[Sec.~2]{OPhd},
which in turn builds on a clustering method of 
Czygrinow, Ha\'{n}\'{c}kowiak, and Wawrzyniak~\cite{CHW08}.
The algorithm is also similar to the ``Binary Bor\r{u}vka'' algorithm~\cite{PR08} 
for finding a minimum-weight spanning tree.  
The global algorithm works iteratively, coarsening the partition in each iteration.
Initially each vertex is in its own part of the partition, and in each iteration some
subsets of parts are merged into larger (connected) parts. The decisions
regarding these merges are based on the numbers of edges between parts, as
well as on certain random choices. Applying the analysis in~\cite{OPhd} it is
possible to show that with high constant probability, after $O(\log(1/\eps))$ iterations,
the number of edges between parts is at most $\eps |V|$, as required.
 Since the sizes of the parts obtained after the last merging step
 may be much larger than desired, as a final step
 it is possible to refine the partition without increasing the number of edges crossing
 between parts by too much by applying an algorithm of Alon Seymour and Thomas~\cite{AST90}.

When turning to the local emulation of the global algorithm by the partition oracle, 
the query complexity and running time of the partition oracle depend on the sizes 
of the parts in the {\em intermediate\/} stages of the algorithm. 
These sizes  are bounded as a function of $1/\eps$ and $d$, but can still
be quite large. As an end-result, the partition oracle described in~\cite{OPhd}
has complexity that grows exponentially with $\poly(1/\eps)$. To reduce this complexity,
we modify the global partition algorithm as follows: If (following a merging stage)
 the size of a part goes above a
certain threshold, we `break' it into smaller parts. To this end we apply to
each large part (in each iteration)
the abovementioned algorithm of Alon Seymour and Thomas~\cite{AST90}. This algorithm 
finds (in graphs with an excluded minor) a relatively small vertex separator whose removal creates
small connected components. Each such refinement of the partition increases the number
of edges crossing between parts. However, we set the parameters for the
algorithm in~\cite{AST90}
so that the decrease in the number of edges between parts due to the 
merging steps dominates the increase due to the `breaking' steps.
One could have hoped that since the sizes of the parts are now always bounded
by some polynomial in $1/\eps$ (and $d$), the complexity of the  partition oracle
will by $\poly(d/\eps)$ as well. However, this is not the case, since in order
to determine the part that a vertex, $v$, belongs to after a certain iteration, it
is necessary to determine the parts that other vertices in the local neighborhood of $v$ belong 
to in previous iterations. This leads to a recursion formula whose solution is quasi-polynomial
in $1/\eps$.

\subsection{Other Related Work}
Yoshida and Ito~\cite{YI10} were the first to provide a testing algorithm for a 
minor-closed property whose complexity is polynomial in $1/\eps$ and $d$. 
\ifnum\icalp=0
Specifically, they
\else
They
\fi
 give a testing algorithm for the property of being outerplanar.
Their result was generalized by Edelman et al.~\cite{EHNO11}
who design a partition oracle with complexity $\poly(d/\eps)$ 
for the class of bounded treewidth graphs.
Known families of graphs with bounded treewidth include cactus graphs, 
outerplanar graphs and series-parallel graphs. 
However, many graphs with an excluded minor do not have bounded treewidth. 
For example, planar graphs are known to have treewidth of $\Omega(\sqrt{n})$. 

Building on the partition oracle of~\cite{HKNO09},
Newman and Sohler~\cite{NS11} design an algorithm for testing any property 
of graphs under the promise that the input graph is taken from $\mathcal{C}$, 
for any $\mathcal{C}$ that is a $\rho$-hyperfinite family of graphs. 
The number of 
queries their algorithm makes to the graph is independent of $|V|$ but is at least exponential in $1/\eps$. 
In a recent work, Onak~\cite{Ona12} proves that there exists a property such that
testing this property requires 
performing $2^{\Omega(1/\eps)}$ queries even under the promise that the input graph is 
taken from a hyperfinite family of graphs. This family of graphs $\mathcal{T}$
consists of graphs that are unions of bounded degree trees.
Onak~\cite{Ona12} defines a subclass of $\mathcal{T}$
and shows that every algorithm for testing the property of
membership in this subclass must perform $2^{\Omega(1/\eps)}$ queries
to the graph.

Czumaj, Shapira, and Sohler~\cite{CSS09} investigated another promise problem. They proved that any hereditary property, namely a property that is closed under vertex removal, can be tested in time independent of the input size if the input graph belongs to a hereditary and non-expanding family of graphs.

As for approximation under a promise, Elek~\cite{Ele10} proved that under the promise that the input graph, $G=(V,E)$, has sub-exponential growth and bounded degree, the size of the minimum vertex cover, the minimum dominating set and the maximum independent set, can be approximated up to an $\eps |V|$-additive error with time complexity that is independent of the graph size. 
Newman and Sohler~\cite{NS11} showed 
how to obtain an $\eps |V|$-additive approximation for a large class of graph parameters.

\section{Preliminaries}
In this section we introduce several definitions and some known results that will be used
in  the following sections.
Unless stated explicitly otherwise, we consider simple graphs, that is, with no
self-loops and no parallel edges. The graphs we consider have a known degree bound $d$,
and we assume we have query access to their incidence-lists representation. 
Namely, for any vertex $v$ and index $1 \leq i \leq d$ it is possible to obtain
the $i^{\rm th}$ neighbor of $v$ (where if $v$ has less than $i$ neighbors, then
a special symbol is returned). If the graph is edge-weighted, then the weight of
the edge is returned as well.

For a graph $G = (V,E)$ and two sets of vertices $V_1,V_2 \subseteq V$, we let
$E(V_1,V_2)$ denote the set of edges in $G$ with one endpoint in $V_1$ and one endpoint in $V_2$.
That is $E(V_1,V_2) \eqdef \{(v_1,v_2)\in E:\; v_1 \in V_1,v_2\in V_2\}$. 
\BD  
\label{def:part-contract}
Let $G=(V, E, w)$ be 
an edge-weighted graph and let 
$\calP = (V_1, \ldots, V_t)$ be a partition of the vertices of $G$ such that 
for every $1 \leq i \leq t$, the subgraph induced by $V_i$ is connected.
Define the {\em contraction\/} $G/ \calP$ of $G$ with respect to the partition 
$\calP$ to be the edge-weighted graph $G' = (V', E', w')$ where:
\BE
\item  $V' = \{V_1, \ldots, V_t\}$ (that is, there is a vertex in $V'$ for
each subset of the partition $\calP$);
\item $(V_i, V_j) \in E'$ if and only if $i \neq j$ and  $E(V_i,V_j) \neq \emptyset$;
\item $w'((V_i, V_j)) = \sum_{(u,v) \in E(V_i,V_j)} w((u,v))$.
\EE
\ED
As a special case of Definition~\ref{def:part-contract} we get the standard
notion of a single-edge contraction.
\BD\label{def:edge-contract}
 Let $G=(V, E, w)$ be an edge-weighted graph on $n$ vertices $v_1, \ldots, v_n$, and
 let $(v_i, v_j)$ be an edge of $G$. The graph obtained from $G$ by {\em contracting}
 the edge $(v_i,v_j)$ is $G/\mathcal{P}$ where $\mathcal{P}$ is the partition of 
 $V$ into $\{v_i, v_j\}$ and singletons $\{v_k\}$ for every $k\neq i, j$. 
 \ED
\BD
For $\eps \in (0,1]$, $k \geq 1$ and a graph $G = (V,E)$,
we say that a partition $\calP = (V_1,\dots,V_t)$ of $V$ is an {\em $(\eps,k)$-partition}
(with respect to $G$), if the following conditions hold:
\BE
\item For every $1 \leq i \leq t$ it holds that $|V_i|\leq k$; 
\item For every $1 \leq i \leq t$ the subgraph induced by $V_i$
in $G$ is connected;
\item The total number of edges whose endpoints are in different parts of the partition
is at most $\eps|V|$ 
(that is, $\left|\left\{(v_i,v_j)\in E:\;v_i\in V_j,v_j\in V_j,i\neq j\right\}\right| \leq \eps|V|$).
\EE
\ED
Let $G=(V,E)$ be a graph and let $\calP$ be a partition of $V$. 
We denote by $g_{\calP}$ the function from $v\in V$ to $2^{V}$ (the set of all subsets of $V$), 
that on input $v\in V$, returns the subset $V_{\ell} \in \calP$ such that $v\in V_{\ell}$.
\BD[\cite{HKNO09}]
An oracle $\mathcal{O}$ is a {\em partition oracle\/} if,
given query access to the incidence-lists representation of a graph
$G=(V,E)$, the oracle $\mathcal{O}$ provides query access to
a partition $\mathcal{P} = (V_1, \ldots, V_t)$ of $V$, 
where $\mathcal{P}$ is determined by $G$ and the internal randomness of the oracle. 
Namely, on input $v\in V$, the oracle returns $g_{\calP}(v)$
and for any sequence of queries, $\mathcal{O}$ answers consistently with the 
\ifnum\icalp=0
same\footnote{While the partition $\calP$ does not depend on the sequence of queries
to the oracle, the oracle may keep in its memory the identity of the
 vertices it was previously queried
on as well as any additional information it has acquired in previous queries and 
the outcome of coins it has flipped.} 
\else
same
\fi
$\calP$. 
An oracle $\mathcal{O}$ is an $(\eps,k)$-{\em partition oracle with respect
to a class of graphs $\mathcal{C}$}
if the partition $\mathcal{P}$ it answers according to has the following properties.
\BE
\item For every $V_{\ell} \in \calP$ , $|V_{\ell}| \leq k$ and the subgraph induced 
by $V_{\ell}$  in $G$ is connected.
\item If $G$ belongs to $\mathcal{C}$, then 
$|\{(u,v)\in E : g_{\calP}(v) \neq g_{\calP}(u)\}| \leq \eps |V|$ with high constant probability,
where the probability is taken over the internal coin flips of $\mathcal{O}$.
\EE
\ED
By the above definition, if $G \in \mathcal{C}$, then with high constant
probability the partition $\calP$ is an $(\eps,k)$-partition, while if $G \notin \mathcal{C}$
then it is only required that each part of the partition is connected and has size at most $k$.
We are interested in partition oracles that have small query complexity, 
namely, that perform few queries to the graph (for each vertex they are queried on).

Recall that a graph $H$ is called a {\em minor} of a graph $G$ if $H$ is isomorphic to a graph that can be obtained by zero or more edge contractions on a subgraph of $G$. A graph $G$ is {\em $H$-minor free} if $H$ is not a minor of $G$.
We next quote two results that will play a central role in this work.
\BF[\cite{OPhd}]\label{fct:forest}
Let $H$ be a fixed graph with $c_1(H)$ edges.
For  every $H$-minor free graph $G=(V,E)$ it holds that:
\begin{newitemize}
\item $|E| \leq c_1(H) \cdot |V|$; 
\item $E$ can be partitioned into at most $c_1(H)$ forests.  
\end{newitemize}
\EF
\BP[\cite{AST90}]
\label{prop:ast90}
Let $G = (V,E)$ be a $K_h$-minor free graph where the vertices of $G$ are associated
with non-negative weights that sum to 1.
There exists a constant $c$ such that
for any $\beta\in(0,1]$, there is a set of  $c h^{3/2}|V|^{1/2}/\beta^{1/2}$
vertices of $G$ whose removal leaves $G$ with no connected component having 
weight greater than $\beta$. Such
a set can be found in time $O(h^{1/2}|V|^{1/2}|E|)$.
\EP 
\ifnum\icalp=1
As a corollary we get (see proof in Appendix~\ref{sec:miss}):
\else
As a corollary we get:
\fi
\BC\label{lm:part}
Let $H$ be a fixed graph. 
There is a constant $c_2(H) > 1$ such that  
for every $\gamma \in (0,1]$, every $H$-minor free graph $G=(V,E)$ with degree bounded by 
$d$ is $(\gamma, c_2(H) d^2/\gamma^2)$-hyperfinite. Furthermore, a $(\gamma, c_2(H) d^2/\gamma^2)$
partition of $V$ can be found in
time $O(|V|^{3/2})$.
\EC
\def\CorProof{
\sloppy
We apply Proposition~\ref{prop:ast90} with equal weights $1/|V|$ to all vertices and
with $\beta = (c^2 h^3 d^2)/(\gamma^2 |V|)$. By Proposition~\ref{prop:ast90}, we obtain a set
$S$ of vertices such that $|S| = (\gamma/d)|V|$ and such that the removal of
$S$ leaves $G$ with connected components of size at most $c^2 h^3 d^2 /\gamma^2 = c_2(H) d^2/\gamma^2$
each. Consider the partition $\calP$ of $V$ that consists of a singleton subset for each vertex in $S$
and a subset for each component (containing the vertices in the component). 
Since the number of edges incident to $S$ is at most $\gamma |V|$, the partition
$\calP$ is a $(\gamma,c_2(H) d^2/\gamma^2)$ partition, and
since $|E| = O(|V|)$ (as the graph is $H$-minor free for a fixed $H$),
the running time for finding $\calP$ is as stated.
}
\ifnum\icalp=0
\OBPF
\CorProof
\OEPF
\fi

\section{A Global Partitioning Algorithm}\label{sec:glob-part}
Our partition oracle is local, in the
sense that its output is determined by the local neighborhood of the vertex it is queried on.
However, as in previous work, the oracle is based 
on a global partitioning algorithm, which accesses the whole graph,
and the oracle emulates this algorithm locally.
In this section we describe this global partition algorithm.
As noted in the introduction, our algorithm and its analysis are based
on~\cite{OPhd} (which in turn builds on a clustering method of
Czygrinow, Ha\'{n}\'{c}kowiak, and Wawrzyniak~\cite{CHW08}, and is also
similar to the ``Binary Bor\r{u}vka'' algorithm~\cite{PR08}
for finding a minimum-weight spanning tree).

The algorithm proceeds in iterations, where in iteration $i$ it
considers a graph $G^{i-1}$, where $G^{i-1}$ is edge-weighted. 
The vertices of $G^{i-1}$ correspond to (disjoint) subsets of vertices that
induce connected subgraphs in $G$, and the
weight of an edge between two vertices in $G^{i-1}$ is the number of edges 
in $G$ between the two corresponding subsets of vertices.
Initially, the underlying graph $G^0$ is $G$ and all edges have weight $1$. 
In each iteration the algorithm contracts a subset of the edges so that
each vertex in $G^i$ corresponds to a subset that is the union of subsets of
vertices that correspond to vertices in $G^{i-1}$. 
When the algorithm terminates it outputs the partition into subsets that
correspond to the vertices of the final graph.

Each iteration of the algorithm consists of two phases. In
the first phase of iteration $i$, a subset of the edges of $G^{i-1}$
 are contracted, resulting in a graph $\widetilde{G}^i$. In the second phase,
some of the subsets that correspond to vertices in $\widetilde{G}^i$
remain as is, and some
are `broken' into smaller subsets. The vertices of $G^i$
correspond to these subsets (both `broken' and `unbroken').
Observe that if the graph $G$ is $H$-minor free for some fixed graph $H$, then
every $G^i$ and $\widetilde{G}^i$ is $H$-minor free as well.

In the first phase of iteration $i$, the contracted edges are selected randomly as follows. 
Each vertex in $G^{i-1}$ selects
an incident edge with maximum weight and tosses a fair coin to be `Heads' or 
`Tails'.  Each selected edge is contracted if and only
if it is selected by a `Heads' vertex and its other endpoint is a `Tails' vertex.
This way, in each iteration, the contracted edges form stars (depth-1 trees).
Therefore, a vertex in the graph $\widetilde{G}^i$ that results from
the contraction of edges in $G^{i-1}$ as described above, corresponds 
to a subset of vertices in $V$ that induces a connected subgraph  in $G$,
and $\widetilde{G}^i = G/\widetilde{\calP}^i$ (recall Definition~\ref{def:part-contract})
where $\widetilde{\calP}^i$ is this partition into subsets.
Since each vertex in each $\widetilde{G}^i$ corresponds to a connected subgraph in
$G$,  we shall refer to the vertices of $\widetilde{G}^i$ 
 as {\em connected components\/} (to be precise, they are connected components
in the graph resulting from removing all edges in $G$ the correspond to (weighted)
edges in $\widetilde{G}^i$).
In each iteration, following the contraction of edges, if the size of
a component goes above a certain 
\ifnum\icalp=0
threshold,\footnote{One may consider 
setting different thresholds for different iterations. However, in our
analysis this does not
seem to give a better bound on the complexity as compared to setting a common 
threshold.} 
\else
threshold,
\fi
$k = \poly(d/\eps)$, then the component is `broken' into smaller connected
components, each of size at most $k$.
This is done using the algorithm referred to in Corollary~\ref{lm:part},
and $G^i$ is the (edge-weighted) graph whose vertices correspond to the 
new components.

\begin{algorithm}[ht!]\label{alg:global}
\caption{A global $(\eps, c_2(H) d^2/\eps^2)$-partition algorithm for an $H$-minor free graph $G=(V,E)$}
\BE
\item Set $G^0 := G$ 
\item For $i = 1$ to $\ell = \Theta(\log 1/\eps)$:
\BE
\item Toss a fair coin for every vertex in $G^{i-1}$.  
\item 
For each vertex $u$ let $(u,v)$ be an edge with maximum weight
that is incident to $u$ (where ties are broken arbitrarily). 
If $u$'s coin toss is `Heads' and $v$'s coin toss is `Tails', then contract $(u,v)$. \label{st:contract} 
\item Let $\widetilde{G}^i = (\widetilde{V}^i, \widetilde{E}^i, \widetilde{w}^i)$ denote 
the graph resulting from the contraction of the edges as determined in the previous step. 
Hence, each vertex $\widetilde{v}_j^i \in \widetilde{V}^i$ corresponds to a subset of vertices 
in $G$, which we denote by $\widetilde{C}^i_j$.
\item Let $\gamma = \eps/(3\ell)$. For each $\widetilde{C}^i_j$ such that 
$|\widetilde{C}^i_j| > c_2(H)/\gamma^2$, partition the vertices in 
$\widetilde{C}^i_j$ into connected subsets of size at most 
$k = c_2(H) d^2/\gamma^2$ each
by running the algorithm referred to in Corollary~\ref{lm:part} on the subgraph induced 
by $\widetilde{C}^i_j$ in $G$.
\label{st:breakup}
\item Set $G^i := G/\mathcal{P}^i$, where $\mathcal{P}^i$ is the 
partition resulting from the previous step. 
\EE
\item For each subset ${C}^{\ell}_j$ in $\mathcal{P}^\ell$
 such that $|{C}^{\ell}_j| > c_2(H)d^2/\eps^2$, 
 partition the vertices in ${C}^{\ell}_j$ into connected subsets each
 of  size at most $3c_2(H)d^2/\eps^2$ by running  the algorithm referred to
 in Corollary~\ref{lm:part}, and output the resulting partition.
\label{st:last}
\EE
\end{algorithm}
\BT\label{thm:global}
Let $H$ be a fixed graph. 
If the input graph
$G$ is $H$-minor free and has degree bounded by $d$, then for any given $\eps \in (0,1]$, 
Algorithm~\ref{alg:global} outputs an $(\eps, O(d^2/\eps^2))$-partition of $G$ 
with high constant probability.
\ET
\OBPF
We first claim that in each iteration, after Step~\ref{st:contract}, 
the total weight of the edges in the graph is decreased by a factor of 
$\left(1-\frac{1}{8c_1(H)}\right)$ with probability at least $\frac{1}{8c_1(H) -1}$,
where 
 the probability is taken over the coin tosses of the algorithm. 
Fixing an iteration $i$, by Fact~\ref{fct:forest} we know that the edges 
of $G^{i-1}$ can be partitioned into at most $c_1(H)$ forests. 
It follows that one of these forests contains edges with total weight at 
least $\frac{w(G^{i-1})}{c_1(H)}$ where $w(G^{i-1})$ denotes the total weight 
of the edges in $G^{i-1}$. 
Suppose we orient the edges of the forest from roots to leaves, 
so that each vertex in the forest has in-degree at most $1$. 
Recall that for each vertex $v$, the edge selected by
$v$ in Step~\ref{st:contract} is the heaviest among its incident edges. 
It follows that the expected total weight of edges contracted in Step~\ref{st:contract} 
is at least $\frac{w(G^{i-1})}{c_1(H)}$ (recall that an edge $(v,u)$
selected by $v$ is contracted if the coin flip of $v$ is `Heads'
and that of $u$ is `Tails', an event that occurs with probability $1/4$).
Thus, the expected total weight of edges that are {\em not\/} contracted is at 
most $w(G^{i-1}) - \frac{w(G^{i-1})}{4c_1(H)} = \left(1-\frac{1}{4c_1(H)}\right)w(G^{i-1})$. 
By Markov's inequality the probability that the total weight of edges that
 are not contracted is at least $\left(1-\frac{1}{8c_1(H)}\right)w(G^{i-1})$ is at most
\ifnum\icalp=0
\BEQ
\frac{\left(1-\frac{1}{4c_1(H)}\right)w(G^{i-1})}
{\left(1-\frac{1}{8c_1(H)}\right)w(G^{i-1})} = 1 - \frac{1}{8c_1(H) -1}\;.\label{eq1}
\EEQ
\else
  $1 - \frac{1}{8c_1(H) -1}$.
\fi
We say that an iteration $i$ is {\em successful\/}
if $w(\widetilde{G}^{i}) \leq \left(1-\frac{1}{8c_1(H)}\right)w(G^{i-1})$.
\ifnum\icalp=0
By Equation~\ref{eq1}, the probability of success is at least $\eta = \frac{1}{8c_1(H) -1}$, for every iteration. 
Let $Y_i$ be the random variable that takes the value $-1/\eta +1$ if the $i$-th iteration is successful and takes the value $1$ otherwise. 
Let $Z_i = \sum_{j=1}^i Y_j$, then $\E(Z_{k+1} | Z_1, \ldots Z_k) \leq Z_k + 1\cdot (1-\eta) + (1-1/\eta)\cdot \eta = Z_k$ and $|Z_k - Z_{k-1}| \leq \frac{1}{\eta}$ for every $k$. By Azuma's inequality~\cite{Azu67} we obtain that $\Pr(Z_{\ell} \geq t) \leq e^{-t^2/(2\ell(1/\eta)^2)}$ for every $t$. Setting $t = 3\sqrt{\ell}/\eta$, we obtain that with probability greater than $9/10$, $Z_\ell < 3\sqrt{\ell}/\eta$.
Let $s$ denote the number of successful iterations, then $Z_\ell = (\ell - s) + s (1-1/\eta) = \ell - s/\eta$. Thus, we obtain that $s > \eta \ell -3\sqrt{\ell}$.
We conclude that with probability at least $9/10$,
 the number of successful iterations is at least $\frac{\ell}{16c_1(H) -2}$.
\else
Using martingale analysis it can be shown that with probability at least $9/10$,
 the number of successful iterations is at least $\frac{\ell}{16c_1(H) -2}$ 
(see Appendix~\ref{sec:miss2} for the full analysis).
\fi

Our second claim is that for every $i$, after Step~\ref{st:breakup}, it holds that 
$w(G^i) \leq w(\widetilde{G}^i) + \frac{\eps  n}{3\ell} $ (where $n= |V|$). This follows from 
Corollary~\ref{lm:part}: For each component $\widetilde{C}^i_j$ that we break, 
we increase the total weight of edges between components
by an additive term of at most $\gamma  |\widetilde{C}^i_j| = \frac{\eps |\widetilde{C}^i_j|}{3\ell}$. 
Thus, after Step~\ref{st:breakup} of the $\ell^{\rm th}$ iteration, with
probability at least $9/10$, the weight of the edges
in $G^\ell$ 
\ifnum\icalp=0
 is at most 
\BEQ
c_1(H)\cdot  n\cdot \left(1-\frac{1}{8c_1(H)}\right)^{\frac{\ell}{16c_1(H) -2}} + 
  \ell \cdot \frac{\eps  n}{3\ell}\leq 2\eps  n /3\;.
\EEQ 
\else
 is upper bounded by $ 2\eps  n /3$.
\fi
Since we add at most $\eps  n/3$ weight in Step~\ref{st:last} (when breaking the
subsets corresponding to vertices in $G^\ell$ into subsets of size at most
$3c_2(H)d^2/\eps^2$), we obtain the desired result.  
\OEPF

\def\FigLocal{
\begin{figure}
\begin{center}
\begin{tikzpicture}[-]
  \tikzstyle{vertex}=[circle,fill=black!25,minimum size=17pt,inner sep=0pt, draw=black]
\tikzstyle{cc}=[circle,fill=white, minimum size=80pt,draw=blue, dashed]
\tikzstyle{cc2}=[circle,fill=black!10, minimum size=80pt,draw=blue, dashed]
\tikzstyle{cc3}=[ellipse,fill=white,draw=black, minimum width= 210pt, minimum height= 100pt, dashed]

  \node[cc2,xshift=6cm,yshift=-2cm] {$C_0$}; 
\foreach \name/\angle/\text in {A-1/234/1, A-2/162/2, 
                                  A-3/90/3, A-4/18/4}
    \node[vertex,xshift=6cm,yshift=-2cm] (\name) at (\angle:1cm) {};

\node[cc,xshift=6cm,yshift=2cm] {$C_1$};  
\foreach \name/\angle/\text in {B-1/234/5, B-2/162/6, 
                                  B-3/90/7, B-4/18/8, B-5/-54/9}
    \node[vertex,xshift=6cm,yshift=2cm] (\name) at (\angle:1cm) {};

\node[cc3,xshift=10.75cm,yshift=0cm] {};
\node[xshift=10.75cm,yshift=1.80cm, above]{$C^{i}(v)$};
    \node[cc2,xshift=9cm,yshift=0cm] {$C^{i-1}(v)$};
  \foreach \name/\angle/\text in {C-1/234/10, C-2/162/11, 
                                  C-3/90/12, C-4/18/13, C-5/-54/14}
    \node[vertex,xshift=9cm,yshift=0cm] (\name) at (\angle:1cm) {};
 \node[vertex,xshift=9cm,yshift=0cm] (C-2) at (162:1cm) {$v$};

    \node[cc,xshift=12.5cm,yshift=0cm] {$C_2$};
 \foreach \name/\angle/\text in {D-1/234/10, D-2/162/11, 
                                  D-3/90/12, D-4/18/13, D-5/-54/14}
    \node[vertex,xshift=12.5cm,yshift=0cm] (\name) at (\angle:1cm) {};

    \node[cc,xshift=15.5cm,yshift=2cm] {$C_4$};
 \foreach \name/\angle/\text in {E-1/234/15, E-2/162/16, 
                                  E-3/90/12, E-4/18/13, E-5/-54/14}
    \node[vertex,xshift=15.5cm,yshift=2cm] (\name) at (\angle:1cm) {};

    \node[cc2,xshift=15.5cm,yshift=-2cm] {$C_3$};
 \foreach \name/\angle/\text in {F-1/234/15, F-2/162/16, 
                                  F-3/90/12}
    \node[vertex,xshift=15.5cm,yshift=-2cm] (\name) at (\angle:1cm) {};

\foreach \from/\to in {1/2,2/3,3/4, 4/5}
    { \draw (C-\from) -- (C-\to); }

  \foreach \from/\to in {1/2,2/3,4/1}
    { \draw (A-\from) -- (A-\to); }

  \foreach \from/\to in {1/2,3/4,4/5,1/3,3/5}
    { \draw (B-\from) -- (B-\to); }

\foreach \from/\to in {1/2,2/3,3/4, 1/5}
    { \draw (D-\from) -- (D-\to); }

\foreach \from/\to in {1/2,2/3, 3/1}
    { \draw (F-\from) -- (F-\to); }

\foreach \from/\to in {1/2,1/3,1/4, 4/5}
    { \draw (E-\from) -- (E-\to); }

  \draw (A-1) .. controls +(-30:2cm) and +(-150:1cm) .. (F-1);
\draw (A-4) -- (C-1);
\draw (B-5) -- (C-2);
\draw (B-4) -- (C-3);
\draw (C-3) -- (D-3);
\draw (C-4) -- (D-2);
\draw (C-5) -- (D-1);
\draw (E-1) -- (F-3);
\draw (E-5) -- (F-3);
\draw (D-5) -- (F-2);
\end{tikzpicture}
\end{center}
\caption{The dark components are `Heads' vertices and the bright ones are `Tails' vertices. Before the $i^{th}$ iteration there are $3$ edges in the cut between $C^{i-1}(v)$ and $C_2$ and $2$ edges in the cut between $C^{i-1}(v)$ and $C_1$. Since $C^{i-1}(v)$ is a `Heads' vertex it merges with $C_2$ which is a `Tails' vertex. $C_3$ does not merge with $C_2$ since the cut $(C_3, C_4)$ is larger than the cut $(C_3, C_2)$. $C_0$ does not merge with any vertex since it is a `Tails' vertex that is not connected to a `Heads' vertex. This illustration is for a case that $C^{i}(v) = \widetilde{C}^{i}(v)$. }
\label{fig-comp}
\end{figure}
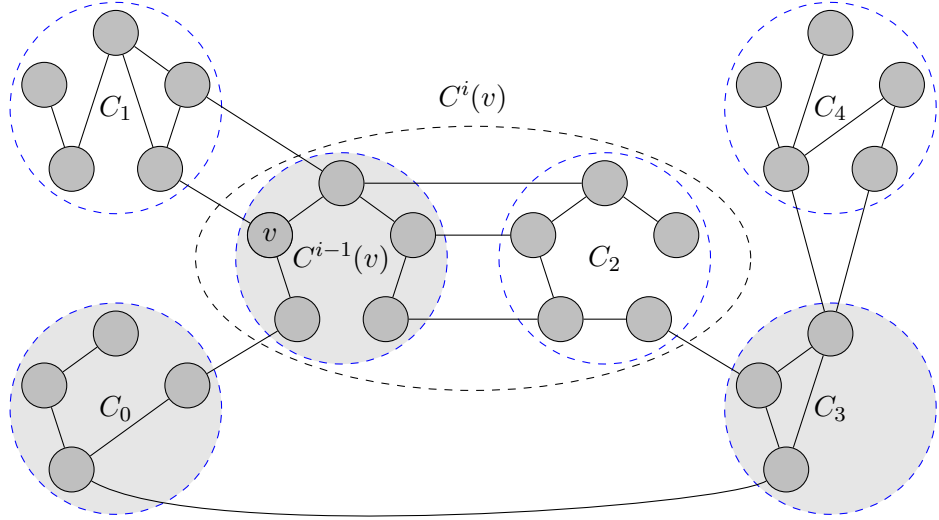
 }
 \ifnum\icalp=0
 \FigLocal
 \fi

\section{The Partition Oracle}
In this section we describe how, given query access to the incidence-lists
representation of a graph $G = (V,E)$ and a vertex $v \in V$, it is possible to emulate 
Algorithm~\ref{alg:global} 
locally and determine the part that $v$ belongs to in the partition
$\calP$ that the  algorithm  outputs. Namely, we prove the following theorem.
\BT\label{tm:local}
For any fixed graph $H$ 
there exists an $(\eps,O(d^2/\eps^2))$-partition-oracle for $H$-minor free graphs that makes 
$(d/\eps)^{O(\log (1/\eps))}$ queries to the graph for each query to the oracle.
The total time complexity of a sequence of $q$ queries to the oracle is $q \log q \cdot (d/\eps)^{O(\log (1/\eps))}$.  
\ET

\OBPF
\sloppy
Recall that the partition $\calP$ is determined randomly
based on the  `Heads'/`Tails' coin-flips of the vertices in each iteration.
\ifnum\icalp=0
Also recall that $g_{\calP}(v)$ denotes the subset of vertices that $v$ belongs
to in $\calP$.
\fi
Since we want the oracle to be efficient, the oracle will flip coins ``on the fly'' as 
required for determining $g_{\calP}(v)$. 
Since the oracle has to be consistent
with the same $\calP$ for any sequence of queries it gets, it will
keep in its memory all the outcomes of the coin flips it has made.
For the sake of simplicity, whenever an outcome of a coin is required, 
we shall say that a coin is flipped, without explicitly stating that
first the oracle checks whether the outcome of this coin flip has
already been determined.
We shall also explain subsequently how to break ties (deterministically) in the choice
of a heaviest incident edge in each iteration of the algorithm.

Recall that the algorithm constructs a sequence of
graphs $G^0=G,$ $\widetilde{G}^1,$ $G^1,$ $\dots,\widetilde{G}^\ell,G^\ell$,
and  that for each $0 \leq i \leq \ell$, the vertices in $G^i$
correspond to connected subgraphs of $G$ (which we refer to as components). 
For a vertex $v\in V$ let  $C^i(v)$ denote the vertex/component that $v$ it belongs to
in $G^i$, and define $\widetilde{C}^i(v)$  analogously with respect to $\widetilde{G}^i$.
Indeed, we shall refer to vertices in $G^i$ ($\widetilde{G}^i$)
 and to the components that correspond to
them, interchangeably.
When the algorithm flips a coin for a vertex $C$ in $G^i$, we may think of the coin flip as
being associated with the vertex having the largest id (according to some arbitrary ordering) in
the corresponding component in $G$.
When the algorithm selects a heaviest edge incident to $C$ and there 
are several edges $(C,C_1),\dots,(C,C_r)$ with the same maximum weight,
it breaks ties by selecting the edge $(C,C_j)$ for which $C_j$ contains
the vertex with the largest id (according to the same abovementioned arbitrary ordering).
We can then refer to $(C,C_j)$ as {\em the\/} heaviest edge incident to $C$.
In particular, since in $G^0$ all edges have the same weight, the heaviest
edge incident to a vertex $u$ in $G^0$ is the edge $(u,y)$ for which $y$ is maximized.

Let $Q^i(v)$ denote the number of queries to $G$ that are performed in order to determine 
$C^i(v)$,  
and let $Q^i$ denote an upper bound on $Q^i(v)$ that
 holds for any vertex $v$. 
We first observe that $Q^1 \leq d^2$. In order to determine $C^1(v)$,
the oracle first flips a coin for $v$. If the outcome is `Tails'
  then the oracle queries the neighbors of $v$. For each neighbor $u$ of $v$
it determines whether $(u,v)$ is the heaviest edge incident to $u$
(by querying all of $u$'s neighbors).
If so, it flips a coin for $u$, and if the outcome is `Heads', then
the edge is contracted (implying that $u \in \widetilde{C}^1(v)$). 
If $v$ is a `Heads' vertex, then it finds its heaviest incident edge, $(v, u)$
by querying all of $v$'s neighbors.
If $u$ is a `Tails' vertex (so that $(v,u)$ is contracted), then the oracle queries all
of $u$ neighbors, and for each neighbor it queries all of its neighbors. By doing so (and
flipping all necessary coins) it can determine which additional edges
$(u,y)$ incident to $u$ 
are contracted (implying for each that  $y\in \widetilde{C}^1(v) = \widetilde{C}^1(u)$).
In both cases (of the outcome of $v$'s coin flip), the number of queries
performed to $G$ is at most $d^2$.
 Recall that a component as constructed above
is `broken' if it contains more than $k = \tilde{O}(d^2/\eps^2)$
 vertices. Since $|\widetilde{C}^1(v)| \leq d+1$ for every $v$, we have that
$C^1(v) = \widetilde{C}^1(v)$.

For general $i > 1$, to determine the connected component that a vertex $v$ belongs to after
 iteration $i$, we do the following. First we determine the component it belongs
 to after iteration $i-1$, namely $C^{i-1}(v)$, at a cost of at most $Q^{i-1}$ queries. 
Note that by the definition of the algorithm, $|C^{i-1}(v)| \leq k$.
\ifnum\icalp=1
We now have two cases (for an illustration see Fig.~\ref{fig-comp} in Appendix~\ref{app:figs}):
\else
We now have two cases (for an illustration see Figure~\ref{fig-comp}):
\fi

\noindent
{\sf Case 1:} $C^{i-1}(v)$ is a `Tails' vertex for iteration $i$. 
In this case we query all edges incident to vertices in 
 $C^{i-1}(v)$, which amounts to at most $d\cdot k$ edges. 
For each endpoint $u$ of such an edge we find $C^{i-1}(u)$.
 For each $C^{i-1}(u)$ that is `Heads' we determine whether its heaviest incident edge
 connects to $C^{i-1}(v)$ and if so the edge is contracted (so that
$C^{i-1}(u) \subset \widetilde{C}^i(v)$). 
To do so, we need, again, to query all the edges incident to vertices in 
 $C^{i-1}(u)$, and for each endpoint $y$ of such an edge we need to find $C^{i-1}(y)$. 
The weight of each edge $(C^{i-1}(u), C^{i-1}(y))$ is $|E(C^{i-1}(u), C^{i-1}(y))|$
(and since all edges incident to vertices in $C^{i-1}(y)$ have been queried,
this weight is determined).
The total number of vertices $x$ for which we need to find $C^{i-1}(x)$ is
upper bounded by $d^2 k^2$, 
and this is also an upper bound on the number of
queries performed in order to determine the identity of these vertices.

\noindent
{\sf Case 2:} $C^{i-1}(v)$ is a `Heads' vertex in iteration $i$.
 In this case we find its heaviest incident edge in $G^{i-1}$, as previously described
for $C^{i-1}(u)$. 
Let $C'$ denote the other endpoint in $G^{i-1}$.
If $C'$ is a `Tails' vertex then  we apply the same procedure to $C'$ as described
in Case 1 for $C^{i-1}(v)$ (that is, in the case that $C^{i-1}(v)$  is a `Tails' vertex in $G^{i-1}$).
The bound on the number of queries performed is also as in Case 1.
In either of the two cases we might need to `break' $\widetilde{C}^{i}(v)$
(in case $|\widetilde{C}^{i}(v)| > k$)
so as to obtain $C^i(v)$.
However, this does not require performing any additional queries to $G$ since all
edges between vertices in $\widetilde{C}^{i}(v)$ are known, 
 and this step only contributes
to the running time of the partition oracle.
We thus get the following recurrence relation for $Q^i$:
$Q^i = d^2\cdot k^2 + d^2 \cdot k^2 \cdot Q^{i-1}$.
Since $k = \poly(d/\eps)$ we get that 
\ifnum\icalp=0
\BEQ
Q^\ell \leq (d \cdot \poly(d/\eps))^{2\ell} = (d/\eps)^{O(\log (1/\eps))}\;,
\EEQ
\else
$Q^\ell \leq (d \cdot \poly(d/\eps))^{2\ell} = (d/\eps)^{O(\log (1/\eps))}$, \fi
as claimed. 

Finally, we turn to the running time.
Let $T^i(v)$ denote the running time for determining $C^i(v)$. 
By the same reasoning as above we have that 
$T^i \leq O(d^2 \cdot k^2) \cdot T^{i-1} + B$ where $B$ is an upper bound on 
the running time of breaking a connected component at each iteration.
From Corollary~\ref{lm:part} we obtain that $B \leq (d \cdot k^2)^{3/2}$. 
Thus, the running time of the oracle is $(d/\eps)^{O(\log (1/\eps))}$ for a single query.
As explained above, for the sake of consistency, the oracle stores its previous coin-flips. By using a balanced search tree to store the coin flips we obtain that the total running time of the oracle for a sequence of $q$ queries is $q \log q\cdot (d/\eps)^{O(\log (1/\eps))}$, as claimed. 
\OEPF 

\vspace{-2ex}
\section{Applications}

\vspace{-0.5ex}
In this section we state the improved complexity for the applications, of the partition oracle,
which are presented in~\cite{HKNO09}. We obtain an improvement either in the query complexity or in the time complexity for all their applications excluding the application of approximating the distance to hereditary properties in which case the improvement we obtain is not asymptotic. 
\ifnum\icalp=1
\begin{newitemize}
\else
\BI
\fi 
\item Hassidim et al.~\cite{HKNO09} show that for any fixed graph $H$ 
there is a testing algorithm for the property
of being $H$-minor free in the bounded-degree model that
performs $O(1/\eps^2)$ queries to $\mathcal{O}$,  
where $\mathcal{O}$ is an $(\eps d/4, k)$-partitioning oracle for the class of $H$-minor free graphs with
degree bounded by $d$, and has $O(dk/\eps + k^3/\eps^6)$ time complexity. 
By using the partition oracle from Theorem~\ref{tm:local} we obtain that the query and time complexity 
of testing  $H$-minor freeness (in the bounded-degree model)
 is improved from $d^{\poly(1/\eps)}$ to $(d/\eps)^{O(\log{1/\eps})}$.
\item Let $\calP$ be a minor-closed property. According to~\cite{RS95b}, 
$\calP$ can be characterized as a finite set of excluded minors. 
Let $S$ denote this set. 
By taking the proximity parameter to be $\eps/|S|$ and applying the testing algorithm for minor-freeness 
on every minor in $S$ we obtain that the  query and time complexity of testing a minor-closed property in the bounded degree model 
is improved from $2^{\poly(|S|/\eps)}$ to $(|S|/\eps)^{O(\log{|S|/\eps})}$. 
In particular this implies a testing algorithm for planarity with complexity
$(1/\eps)^{O(\log(1/\eps))}$.
\ifnum\icalp=1
\end{newitemize}
\else
\EI
\fi
The next approximation algorithms work under the promise that the input graph is a graph 
with an excluded minor (of constant size). 
Under this promise we obtain the following improvements in the query complexity while the 
time complexity remains unchanged (the former time complexity dominates the improvement in the 
time complexity of the partition oracle):
\ifnum\icalp=1
\begin{newitemize}
\else
\BI
\fi 
\item Hassidim et al.~\cite{HKNO09} provide a constant time $\eps|V|$-additive-approximation algorithm for 
 minimum vertex cover size, maximum independent set size,
and the minimum dominating set size for any family of graphs with an efficient partition oracle.  
The algorithms makes $O(1/\eps^2)$ queries to the partition oracle.
By using the partition oracle from Theorem~\ref{tm:local}, the query complexity of the approximation 
algorithms is improved from $d^{\poly(1/\eps)}$ to $(d/\eps)^{O(\log{1/\eps})}$ 

\item By Lemma 11 in~\cite{HKNO09}, for any finite set of connected graphs $\mathcal{H}$, 
there is an $\eps$-additive-approximation algorithm for the distance to the property of not having any graph 
in $\mathcal{H}$ as an induced subgraph, which makes $O(1/\eps^2)$ queries to the partition oracle. 
Hence, the query complexity of the algorithm is improved from $d^{\poly(1/\eps)}$ to $(d/\eps)^{O(\log{1/\eps})}$.
\ifnum\icalp=1
\end{newitemize}
\else
\EI
\fi
\newcommand{\etalchar}[1]{$^{#1}$}

\ifnum\icalp=1
\newpage
\appendix
\section{Figures}\label{app:figs}

\FigLocal

\section{Missing Details}

\subsection{Proof of Corollary~\ref{lm:part}}\label{sec:miss}
\OBPF
\medskip \CorProof
\OEPF

\subsection{Lower bound on the number of successful iterations}\label{sec:miss2}
Recall that for every iteration the probability of success is at least $\eta = \frac{1}{8c_1(H) -1}$. 
Let $Y_i$ be the random variable that takes the value $-1/\eta +1$ if the $i$-th iteration is successful and takes the value $1$ otherwise. 
Let $Z_i = \sum_{j=1}^i Y_j$, then $\E(Z_{k+1} | Z_1, \ldots Z_k) \leq Z_k + 1\cdot (1-\eta) + (1-1/\eta)\cdot \eta = Z_k$ and $|Z_k - Z_{k-1}| \leq \frac{1}{\eta}$ for every $k$. By Azuma's inequality we obtain that $\Pr(Z_{\ell} \geq t) \leq e^{-t^2/(2\ell(1/\eta)^2)}$ for every $t$. Setting $t = 3\sqrt{\ell}/\eta$, we obtain that with probability greater than $9/10$, $Z_\ell < 3\sqrt{\ell}/\eta$.
Let $s$ denote the number of successful iterations, then $Z_\ell = (\ell - s) + s (1-1/\eta) = \ell - s/\eta$. Thus, we obtain that $s > \eta \ell -3\sqrt{\ell}$.
We conclude that with probability at least $9/10$,
 the number of successful iterations is at least $\frac{\ell}{16c_1(H) -2}$.
\fi
\end{document}